\documentclass[prd, aps, superscriptaddress, preprintnumbers, twocolumn, floatfix, nofootinbib]{revtex4}
\pdfoutput=1

\usepackage{amsfonts}
\usepackage{amsmath}
\usepackage{amssymb}
\usepackage{bm}
\usepackage{dcolumn}
\usepackage{graphicx}   
\usepackage[latin1]{inputenc}
\usepackage{latexsym}
\usepackage{rotating}
\usepackage{hyperref}
\usepackage{graphicx}
\usepackage{color}
\usepackage{cases}

\newcommand\be{\begin{equation}}
\newcommand\ba{\begin{eqnarray}}
\newcommand\ee{\end{equation}}
\newcommand\ea{\end{eqnarray}}

\newcommand{\gsim}{\mathrel{\hbox{\rlap{\lower.55ex \hbox {$\sim$}}
                   \kern-.3em \raise.4ex \hbox{$>$}}}}
\newcommand{\lsim}{\mathrel{\hbox{\rlap{\lower.55ex \hbox {$\sim$}}
                   \kern-.3em \raise.4ex \hbox{$<$}}}}
\newcommand{\red}{\textcolor{red}}
\newcommand{\blue}{\textcolor{blue}}

\begin{document}

\title {Cosmological Magnetic Fields from Ultralight Dark Matter}

\author{Robert Brandenberger}
\email{rhb@physics.mcgill.ca}
\affiliation{Department of Physics, McGill University, Montr\'{e}al, QC, H3A 2T8, Canada} 

\author{J\"{u}rg Fr\"{o}hlich}
\email{juerg@phys.ethz.ch}
\affiliation{Institute of Theoretical Physics, ETH Z\"urich, CH-8093 Z\"urich, Switzerland}

\author{Hao Jiao}
\email{hao.jiao@mail.mcgill.ca}
\affiliation{Department of Physics, McGill University, Montr\'{e}al, QC, H3A 2T8, Canada} 

\date{\today}


\begin{abstract}

We propose a mechanism for the generation of magnetic fields on cosmological scales that is operative after recombination. An essential 
ingredient is an instability (of parametric resonance type) of the electromagnetic field driven by an oscillating pseudo-scalar dark matter field, $\phi$, that is coupled to the electromagnetic field tensor via a $\phi F \wedge F$ term in the Lagrangian of axion-electrodynamics. We find that magnetic fields larger than the observational lower bounds can be generated soon after recombination on scales of $1 {\rm{Mpc}}$.

\end{abstract}

\maketitle

\section{Introduction} 
\label{sec:intro}
 
There is increasing evidence \footnote{See e.g. \cite{Kunze, Widrow2, Durrer, Tanmay} for reviews on cosmological magnetic fields.} 
for the existence of magnetic fields on cosmological scales. Since such fields are present in voids, it is natural to assume that they 
are primordial in origin. In particular, there is a lower bound \cite{Taylor} on the amplitude of inter-galactic magnetic fields from the 
non-observation of $\gamma$-ray cascade emission
 \be
 B(r) \, > \, 10^{-17} {\rm{Gauss}} 
 \ee
 on length scales, $r$, larger than $10^{-1} {\rm{Mpc}}$.\footnote{There is also an upper bound on the strength of the magnetic field of 
 $B < 10^{-9} {\rm{Gauss}}$ on these scales derived from the non-observation of magnetic field signals in the cosmic microwave 
 background \cite{CMB}.} The lower bound on the magnetic field increases on smaller scales, but on such scales the origin of the 
 magnetic fields may be astrophysical. We will not consider them in this letter. 
 
The origin of cosmological magnetic fields remains a mystery.  One popular mechanism for the generation of such fields relies on processes
breaking the scale-invariance in the electromagnetic sector that occur during a primordial period of cosmological inflation (see e.g. \cite{Widrow, 
Ratra}). They may occur if the inflaton,  $\phi_I$,  is a pseudo-scalar field coupling to the photon field via a  $\phi_I F \wedge F$ term (see
\cite{Jackiw, Field1,  Field2, JF1, JF2, Joyce}).\footnote{This mechanism has been explored in the context of inflation in many works (see, e.g., \cite{Sorbo, Durrer2}.) A potential parametric resonance instability was discussed in \cite{Namba, Adshead}, and in \cite{Campa, Durrer2, Caprini} 
it was emphasized that the resulting magnetic field will be helical.} In the presence of such terms in the Lagrangian violating scale-invariance 
it is also possible that magnetic fields are generated in cosmological phase transitions \cite{pt}. A key challenge for the viability of this 
particular mechanism is to show that magnetic fields are generated on sufficiently large scales, because the typical length scale on 
which the magnetic field is produced initially is smaller than the Hubble radius at the time of the phase transition. The fact that, in magnetohydrodynamics, an inverse energy cascade (flow of energy from shorter to longer wavelength modes) has been established 
may help the argument. However, in both mechanisms, one faces the problem that the amplitude of the magnetic field decreases 
between the time of the phase transition and the present.
 
In this letter we propose a mechanism for the generation of magnetic fields on cosmological scales that becomes relevant after the time 
of recombination. Our mechanism links magnetogenesis to the dark matter mystery. We assume that dark matter is ``wave-like'' 
(see \cite{Elisa, Hui} for reviews of wave dark matter) and is described by a pseudo-scalar axion field $\phi$ that couples to electromagnetism 
via the standard $\phi F \wedge F$ term. We assume that dark matter is created via a standard misalignment mechanism and, hence, that the
axion field $\phi$ starts to oscillate coherently in our Hubble patch once the Hubble rate $H$ has fallen below the 
dark-matter mass, $m$. The oscillations of the field $\phi$ induce a ``tachyonic'' resonance in the electromagnetic field, a phenomenon 
previously explored in different contexts, such as  \cite{Evan1} (axion monodromy inflation), \cite{Evan2} (inflationary magnetogenesis), 
\cite{Rudnei} (graviton-induced ALP decay), and \cite{Chunshan} (graviton to photon conversion). In our scenario, magnetic fields are
generated at rather late times, namely after the time of recombination.\footnote{See also the parametric resonance instability to matter 
generation from an oscillating inflaton field at the end of the period of inflation \cite{TB, DK}.}

\section{Magnetic Field Generation from an Oscillating Ultralight Dark Matter Field}

As announced, we assume that dark matter comes from a pseudo-scalar, ultralight dark-matter field $\phi$ with a potential, $V$, that is
quadratic in $\phi$ near its minimum,
\be \label{V}
V(\phi) \, \simeq \, \frac{1}{2} m^{2}\,\phi^2 \, .
\ee
Considerations of structure formation imply a lower bound \cite{Lower} on $m$ of the order of $m > 10^{-20} {\rm{eV}}$ (for lower values of $m$, 
structures on scales probed by Lyman $\alpha$ emitters would be suppressed). Assuming dark matter to be wavelike yields an upper bound on
$m$, namely $m < 10 {\rm{eV}}$; (see e.g. \cite{Rodd}).  The field quanta of $\phi$ can be identified with axion- or axion-like particles 
constituting dark matter \cite{Witten}. Axion fields are ubiquitous in string theory, and it is not hard to argue for the existence of fields 
with ultralight masses (see e.g. \cite{BBF} for a recent construction). 

We will introduce a dimensionless measure of the dark matter mass via
\be
m \, \equiv \, m_{20} 10^{-20} {\rm{eV}} \, .
\ee

A homogeneous coherent scalar-field configuration will remain frozen by Hubble damping until the time when $H \sim m$, where 
$H$ is the Hubble expansion rate. Making use of the Friedmann equation we find that this is the case up to the time when the temperature 
$T$ of the universe drops below a value of roughly 
\be
T \, \sim \, m_{20}^{1/2} 10^4 T_{eq} \, ,
\ee
where $T_{eq}$ is the temperature at the time of matter-radiation equality.  For our mechanism to be relevant for an explanation of the 
observed cosmic magnetic fields, we must require the coherence of the $\phi$ configuration to persist until the time of recombination. 
This implies that one must assume the mass $m$ to be very small, since, in this case, the field $\phi$ starts to oscillate at a rather late
time and, as a consequence, there is no time for its coherence to decay too early.

Axion fields typically interact with the electromagnetic field as described by a Lagrangian
\be \label{int}
{\cal{L}} \, = \, \frac12 \partial^{\mu}\phi \partial_{\mu}\phi - V(\phi) -\frac14 F_{\mu\nu}F^{\mu\nu} + g_{\phi\gamma}\phi F_{\mu\nu}\widetilde{F}^{\mu\nu},
\ee
where $F_{\mu \nu}$ is the field strength tensor of the vector potential $A_{\mu}$, and $\widetilde{F}_{\mu \nu}$ is its dual;  
the coupling constant $g_{\phi \gamma}$ has inverse mass units. Using the homogeneous Maxwell equations to eliminate the
electric field, $\vec{E}$, and assuming that $\phi$ only depends on physical time $t$, one finds
that the Lagrangian in \eqref{int} yields the following equation for the magnetic induction, $\vec{B}$,
\begin{equation} \label{FE}
-\Delta \vec{B} + \ddot{\vec{B}} +3H\,\dot{\vec{B}} + \frac{3}{2}\dot{H} \vec{B} +
(\frac{3}{2} H)^{2} \vec{B} - g_{\phi\gamma}\, \dot{\phi}\vec{\nabla}\wedge \vec{B} = 0,
\end{equation}
where $H=\dot{a}/a$ is the Hubble parameter, $a$ is the scale factor of the background space-time, and the overdot indicates a derivatice 
with respect to physical time $t$, while the field equation for $\phi$, with $V(\phi)$ as in \eqref{V}, is given by
\begin{equation} \label{AX}
\square\, \phi = - g_{\phi\gamma}\, \vec{E}\cdot \vec{B} - m^{2}\phi\,.
\end{equation}
 From the non-observation of events caused by interactions of dark matter with photons there are mass-dependent upper limits on the 
 coupling constant $g_{\phi \gamma}$. A conservative upper bound is (see e.g.  \cite{Marsch} for a review and \cite{site} for a 
 web site with updated bounds)
\be
{\tilde{g}}_{\phi \gamma} \, \ll \, 1 \, ,
\ee
where
\be
g_{\phi \gamma} \, \equiv \, {\tilde{g}}_{\phi \gamma} 10^{-10} {\rm{GeV}}^{-1} \, .
\ee

The interaction term in (\ref{int}) violates CP symmetry and affects the two photon polarization states differently. Thus, it may lead to 
the generation of a helical magnetic field.

Assuming that the axion field $\phi$ only depends on time, the field equations \eqref{FE} yield the following equations of motion 
for the amplitudes, ${\cal{A}}_{\pm}$, of the electromagnetic vector potential in Fourier space (see \cite{Durrer} for a detailed discussion)  
\be \label{EoM}
\bigl( \partial_{\eta}^2 + k^2 \pm k g_{\phi \gamma} a(\eta) {\dot{\phi}} \bigr) {\cal{A}}_{\pm} \, = \, 0 \, ,
\ee
where $\eta$ is conformal time.
 Note that, in an expanding background, $k$ is the comoving wave number. The equations \eqref{FE} and \eqref{EoM} 
are valid in the absence of an electromagnetic plasma, i.e., only during a period when photons propagate without scattering. Thus, they 
are applicable after recombination.

From (\ref{EoM}) we conclude that, in the presence of a dynamical axion field, there is a ``tachyonic'' instability in the evolution of ${\cal{A}}_{\pm}$. 
Depending on the sign of ${\dot{\phi}}$ this instability affects different polarization modes. Under our assumptions, $\phi$ is oscillating in time,  
and in this case during half of the oscillation period one polarization state is unstable, while during the other half of the period the other 
state is unstable. Since $a$ varies in time, the Floquet exponent describing the exponential instability varies in time, and thus, when 
integrated over time, an asymmetry between the two polarization states arises, resulting in a helical magnetic field. 

The instability described here affects only long wavelength modes.  For an oscillating $\phi$ background, ${\dot{\phi}} \sim \, \phi_0 m$, 
with $\phi_0$ the initial amplitude of oscillation at recombination, the critical comoving wavenumber, $k_c$, above which the instability
 shuts off is given by
\be \label{critical}
k_c(\eta) \, \simeq \, g_{\phi \gamma} m \phi_0 a (\eta) \, . 
\ee
Note that the instability shifts to higher values of $k$ as time increases.  Since the instability sets in at the time of recombination 
and proceeds rapidly, we focus on this particular time.

We assume that the field $\phi$ accounts for most of the dark matter in the universe. At the time of recombination $m \phi_0$ must then 
be given by
\be \label{amplitude}
m \phi_0 \, \sim \, T_{{rec}}^2 \, ,
\ee
where $T_{rec} \sim 10^{-1} {\rm{eV}}$ is the temperature of the universe at the time of recombination.  Hence, the comoving 
wavelength, $\lambda_c$, below which the instability is shut off is given by
\be
\lambda_c \, \sim {\tilde{g}}_{\phi \gamma}^{-1} 10^{-6} {\rm{Mpc}} \, .
\ee

Next, we provide an order of magnitude estimate of the magnetic field generated according to our mechanism. The instability in the
evolution of the electromagnetic field manifests itself in an exponential growth of ${\cal{A}}_{\pm}(k)$ with the Floquet exponent, $\mu_k$,
given by
\be \label{eff}
\mu_k \, = \, \bigl( g_{\phi \gamma} a\, m \phi(t) k \bigr)^{1/2}\, ,
\ee
where $t$ is physical time, and $\phi(t)$ here denotes the amplitude of the envelope of the oscillating scalar field which is equal to $\phi_0$ at the time of recombination.  Note that the Floquet exponent for the resonance increases with $k$ until $k = k_c$. Since the phase space of Fourier modes scales as 
$k^3$, most of the dark matter energy flows into modes with $k \sim k_c$.  Evaluated at this value we obtain 
\be \label{eff}
\mu_{k_c} \, = \,  g_{\phi \gamma} a\, m \phi(t)\, .
\ee
The expansion of space is negligible, provided that 
$\mu_k(t) > H(t)$.  Taking into account the decay of the amplitude of $\phi$, namely $\phi(t) \sim T^{3/2}$, where $T$ is the temperature
at time $t$, and making use of the Friedmann equations to determine $H(t)$, the efficiency criterion at time $t$ becomes
\be
g_{\phi \gamma} m \phi_0 \, > \, m_{pl}^{-1} T_{rec}^2 \bigl( \frac{T}{T_{rec}} \bigr)^{3/2} \bigl( \frac{T_{rec}}{T_0} \bigr)
\ee
where $T_0$ is the current temperature of the CMB. With relation (\ref{amplitude}) we obtain the condition
\be \label{cond}
{\tilde{g}}_{\phi \gamma}  \, > \, 10^{-6} \bigl( \frac{T}{T_{rec}} \bigr)^{3/2}
\ee
for efficient resonance.   Note that since $\mu$ decreases in time less fast than $H$, it is possible that the efficiency condition is not satisfied at $T = T_{rec}$, but becomes satisfied later on.

Back-reaction will shut off the resonant production of photons once a fraction ${\cal{F}} \ll 1$ of the dark matter density has been drained.  This shutoff is expected to happen shortly after the time of recombination.
The power spectrum, $P_B(k_c)$ of the magnetic field (which 
is the mean-square value of $B$ in a sphere of radius $k^{-1}$) at the critical scale $k_c$ is thus given by
\be
P_B(k_c) \sim \, {\cal{F}} T_{rec}^4 
\ee
which corresponds to a characteristic position space value of $B$ on the scale $k_c$ of
\be \label{maximal}
B(k_c) \, \sim \, {\cal{F}}^{1/2} {\rm{Gauss}} \, .
\ee
Note that this is the amplitude at $T = T_{rec}$, the time when the instability takes place. Afterwards, the amplitude of $B$ will decay as $a(t)^{-2}$.

The scaling law of the magnetic field on larger length scales follows the one of the magnetic field in a phase transition, which scales with 
wavenumber on length scales larger than the characteristic wavelength of the transition. It is thus given by (see \cite{Durrer} for a review)
\be
B(k) \, \simeq \, \bigl( \frac{k}{k_c} \bigr)^n B(k_c) \, ,
\ee
where $n = 3/2$ for the helical magnetic field component, and $n = 1$ for the non-helical component. Evaluating the result for $n = 3/2$, 
and making use of the field amplitude (\ref{maximal}) at the critical scale and the value (\ref{critical}) for the critical wavenumber, 
the magnetic field on one Megaparsec scale ($k = k_1 \simeq 10^{-38} {\rm{GeV}}$) is found to be given by
\be
B(k_1) \, \sim \, {\tilde{g}}_{\phi \gamma}^{-3/2} {\cal{F}}^{1/2} 10^{-15}\, {\rm{Gauss}} \, ,
\ee
where we have included the dilution after energy transfer.

We conclude that the mechanism described here can give rise to the growth of helical magnetic fields on Megaparsec scales sufficiently large
to be compatible with observational bounds.  Note that the scaling of our final result, $\propto {\tilde{g}}_{\phi \gamma}^{-3/2}$, follows 
from the fact that $k_c$ scales as ${\tilde{g}}_{\phi \gamma}$. This scaling relation only holds down to a value of 
${\tilde{g}}_{\phi \gamma}$ when (\ref{cond}) ceases to be satisfied.  

\section{Conclusions and Discussion}

In this letter we have proposed a plausible mechanism for the generation of magnetic fields on cosmological scales. Our mechanism 
works at late times -- specifically after the time of recombination. It is based on the ``tachyonic'' instability of infrared modes of the 
electromagnetic field, which sets in if a coherently oscillating pseudo-scalar field $\phi$ is coupled to electromagnetism through a 
$\phi F \wedge F$ term in the Lagrangian. This term is typical for an axion field $\phi$, which we have assumed to have an ultralight mass.  

A key point of our proposal is that it provides a link between dark matter and cosmological magnetic fields. A mechanism similar to ours
has been used recently \cite{JH} to provide sufficient Lyman-Werner radiation to allow DCBH (direct collapse black hole) formation 
from energy density fluctuations in the standard cosmological scenario.

\section*{Acknowledgements}

\noindent  This research is supported in part by funds from NSERC and from the Canada Research Chair program.  
JH acknowledges support from a Milton Leung Fellowship in Science.  RB thanks the Institute of Particle and 
Astrophyiscs and the Institute of Theoretical Physics of the ETH for hospitality.

\end{document}